\documentclass[review]{elsarticle}
\usepackage{extarrows, color, makecell}
\usepackage{lineno,hyperref, geometry, setspace, amsmath, amssymb, algorithmic, graphicx, textcomp, amsfonts}
\modulolinenumbers[0]
\journal{Journal of \LaTeX\ Templates}

\begin{document}
\begin{spacing}{1.5}  %%行间距变为1.5倍行距

\begin{frontmatter}

\title{Multi-party quantum privacy comparison of size based on d-level GHZ states}

\author[mymainaddress,mysecondaryaddress]{Hao Cao}
\author[mymainaddress]{Wenping Ma}
\author[mymainaddress]{Liangdong Lyu}
\author[mythirdlyaddress]{Yefeng He}
\author[mymainaddress]{Ge Liu}
\cortext[mycorrespondingauthor]{Hao Cao, caohao2000854@163.com.}

\address[mymainaddress]{State Key Laboratory of Integrated Services Networks, Xidian University, Xi'an, 710071, China}
\address[mysecondaryaddress]{School of Information and Network Engineering, Anhui Science and Technology University, Chuzhou, 233100, China}
\address[mythirdlyaddress]{School of Telecommunications and Information Engineering, Xi'an University of Posts and Telecommunications, Xi'an 710121, China}

\begin{abstract}
Quantum privacy comparison(QPC) plays an important role in secret ballot elections, private auctions and so on. To date, many multi-party QPC(MQPC) protocols have been proposed to compare the equality of $k(k\geq 3)$ participants. However, there are few examples of MQPC used to compare the sizes or values of their privacies. In this paper, we propose a MQPC protocol by which any $k(k\geq 3)$ participants can compare the sizes of their privacies with executing the protocol just once. The proposed MQPC protocol takes the $d-level$ GHZ states as  quantum resources, and a semi-honest $TP$ is introduced to help the participants to determine the relationship of their privacies. Further more, only single-particle unitary transformations and measurements are involved, and the participants need not to share common secrets with each other beforehand which makes the proposed protocol much more efficient. Analysis shows that our protocol is secure against internal and external attack in theory.

\end{abstract}

\begin{keyword}
\texttt{Quantum Cryptography; Multi-party Quantum Privacy Comparison(MQPC); Quantum Fourier Transform(QFT);  Third Party(TP); GHZ State.}
\end{keyword}

\end{frontmatter}

\section{Introduction}
Privacy comparison originates from the concept of millionaire problem introduced by Yao which can be described as follows: two millionaires want to know who is richer without divulging any information about their wealth, and a novel solution for the problem was proposed by him\cite{Y1982}. Later, many solutions\cite{IG2003,LT2005} have been proposed, and the millionaire problem, especially privacy comparison became an important topic in classical cryptography. In the other hand, with the revolutionary application, known as BB84, of the quantum mechanics in the cryptography \cite{BB1984}, quantum cryptography attracts much more attention from all over the world, and many kinds of cryptography protocols such as quantum key distribution (QKD)\cite{LC1999,LMC2005}, quantum secret sharing (QSS)\cite{HBB1999,BLL2017,CM2017-2}, quantum direct communication(QDC)\cite{ZDS2017}, quantum key agreement(QKA)\cite{CM2017-1}, and so on, have been proposed. As an important topic, privacy comparison in the quantum circumstances, i.e., quantum privacy comparison(QPC), has attracted wide attention from many cryptographers.

In 2009, the first two-party QPC protocol for comparing information of equality based on bell states and a hash function was proposed by Yang and Wen \cite{YW2009}. Thereafter, several two-party QPC protocols\cite{CXN2010,XCW2012,TLH2012} based on entangled quantum resources, such as GHZ states, $\chi$-type states and so on, were proposed. However, only two parties were involved in the above protocols. Until 2013, the first multi-party QPC(MQPC) was proposed by Chang et al\cite{CTH2013}. Since then, various two-party \cite{ZLW2014,SYW2015,LXH2017,H2017,XZ2017} and multi-party \cite{WSH2014,HHG2016,Y2016,HHH2017} structures were proposed. However, the aforementioned protocols are only suitable for comparing the equality of information. When it comes to size comparison, these protocols are powerless.

Fortunately, in 2011, Jia et al presented the first two-party QPC protocol for comparing the sizes of privacies based on $d-level$ three-particle GHZ states \cite{JWS2011}, in which the information of sizes was encoded  into the phase of GHZ states. Later, in 2013, three two-party QPC protocols  \cite{LSL2013,GGQ2013,ZLZ2013}  for comparing the information of sizes based on $d-level$ bell states were proposed. In the same year, Yu et al \cite{YGL2013} proposed another one based on  $d-level$ single particles. However, the five QPC protocols mentioned above only relate to comparing the size of two parties. Until 2014, the first protocol of size comparison in multi-party circumstance\cite{LYS2014} was proposed by Luo et al. However, the participants needed to  share a privacy key $K$ beforehand by using QKA protocol which will waste a lot of quantum resources. Besides, each participant and $TP$ need to establish an authenticated classical channel beforehand. Later, Huang et al \cite{HHG2015} proposed another MQPC protocol based on GHZ states, which can also be used to compare the sizes of all privacies. However,we found that there exists a serious security flaw in the protocol after close analysis, i.e.,  an internal participant can get the privacy of any other participant without being found.

In this paper, we will propose a novel MQPC protocol by which any $k(k\geq 3)$ participants can compare the sizes of their privacies with executing the protocol just once. In this protocol, a semi-honest third party($TP$)\cite{LYS2014} is introduced to help the participants to compare the sizes of their information. The semi-honest means that the $TP$ will always execute the protocol honestly, record the information of the participants and try to extract their privacies from the records, but he will not conspire with any participant or outside eavesdropper. First, $TP$ prepares some $d-level$ $k-particle$ GHZ states and distributes them to every participant. Second, each participant encodes them with unitary operations based on a random sequence, and sends them back to $TP$. Next, each participant encrypts his size by the random sequence and sends it to $TP$. At last, $TP$ measures the GHZ states on the Z-basis separately, compares them with the encrypted size, and obtains the results of comparison. The proposed MQPC protocol can ensure that \\
(1)correctness: all participants can get the size relationship of their privacy correctly with the help of $TP$ if they  execute the protocol honestly. \\
(2)security: the semi-honest party $TP$ cannot get any information about the privacies of participants except the size relationship. Besides, each participant cannot deduce privacy of others from the comparison result. \\
The structure of our paper is as follows. Section 2 devotes to the details and correctness of our proposed protocol, and a novel example is presented. Section 3 analyzes the proposed protocol and compares it to the existed protocols, and a brief conclusion is given in section 4.\\
\section{Results}
Before going further, firstly we recall some definitions and quantum resources which will be used in the description of our protocol.
\subsection{Preparation for the protocol}
 The quantum resource used in our protocol is the $d-level$ $k-particle$ GHZ state which can be represented as
\begin{equation}\label{GHZ state}
\begin{array}{l}
|\Phi\rangle=\frac{1}{\sqrt{d}}(\underbrace{|0\rangle|0\rangle\cdots |0\rangle}_k+\underbrace{|1\rangle|1\rangle\cdots |1\rangle}_k+\cdots +\underbrace{|d-1\rangle|d-1\rangle\cdots |d-1\rangle}_k)
\end{array}
 \end{equation}
 For a $d-level$ quantum system, there are two indistinguishable orthogonal bases, Z-basis and X-basis :
\begin{equation}\label{X-basis and Z-basis}
\begin{array}{l}
  Z=\{|0\rangle, |1\rangle, |2\rangle, \ldots, |d-1\rangle\} \\
  X=\{QFT|0\rangle, QFT|1\rangle, QFT|2\rangle, \ldots, QFT|d-1\rangle\}
\end{array}
\end{equation}
 where $QFT: |x\rangle \rightarrow \frac{1}{\sqrt{d}}\displaystyle\sum_{z=0}^{d-1}exp(\frac{2\pi ixz}{d})|x\rangle$ is the quantum Fourier transform(QFT).  Let us introduce an unitary operation (we call it shift operator) as follows:
\begin{equation}\label{Shift operator}
\begin{array}{l}
  U_r = \displaystyle\sum_{t=0}^{d-1}exp(\frac{2\pi it(t\oplus r)}{d})|t\oplus r\rangle \langle t|
\end{array}
\end{equation}
 Hereafter, the symbols $\oplus$ and $\ominus$ denote modular $d$ addition and subtraction. It is easy to verify that the shift operator is an one-to-one map from Z-basis to itself and X-basis to itself, i.e.,
\begin{equation}\label{One-to-one map}
\begin{array}{l}
  U_r (|s\rangle) = |s\oplus r\rangle \\
  U_r (QFT|s\rangle) = QFT|s\oplus r\rangle    (s=0,1,2,\ldots, d-1)
\end{array}
 \end{equation}

\subsection{The MQPC protocol for comparing the sizes of information}

Let $TP$ be a semi-honest party and $P_0, P_1, P_2,\ldots, P_{k-1}$ be $k$ participants. Each participant $P_i$ $( i\in $\{0, 1, 2, $\ldots, k-1\} )$ possesses a $m-length$ privacy $p_i=(p_{i,1}, p_{i,2}, \ldots, p_{i,m})\in \{0, 1, \ldots , l\}^m$ (here $d=2l+1$). They want to compare the size of $p_{0,j}, p_{1,j}, \ldots, p_{k-1,j} (j=1, 2, \ldots, m)$ without revealing any information. Through executing the following protocol, they could achieve their goals with the help of $TP$.  The detailed
description of our MQPC protocol can be seen as follows:

 \textbf{Step 1 Preparation.}  $TP$ prepares $m$ identical $d-level$ $k-particle$ GHZ states in the form of equation ( \ref{GHZ state} ), and splits them into $k$ particle-sequences: $S_0, S_1, \ldots, S_{k-1}$ . The $i-th$ sequence $S_i(i=0, 1, \ldots, k-1)$ is consisting of the $i-th$ particles of these GHZ states. Next, he will get a series of new sequence $S_0^\prime, S_1^\prime, \ldots, S_{k-1}^\prime$  by inserting $m$ decoy particles which are selected from X-basis or Z-basis ( see equation (\ref{X-basis and Z-basis}) ) randomly into each sequence $S_i$, and sends the resulted sequence $S_i^\prime(i=0, 1, \ldots, k-1)$ to the $i-th$ participant $P_i$.

\textbf{Step 2 Eavesdropping Checking.}  After confirming that each participant $P_i$ has received the sequence $S_i^\prime$, $TP$ publishes the position and measurement basis(X-basis or Z-basis) of each decoy particle in $S_i^\prime$. $P_i$ and $TP$ execute eavesdropping checking similar to BB84. If the safety of the channel is not acceptable, the protocol goes to \textbf{Step 1.} Otherwise, the protocol will continue. After successfully passed the eavesdropping checking, each participant $P_i$ will recover the sequence $S_i$ by deleting the decoy particles from $S_i^\prime$ .

\textbf{Step 3 Encoding.}  Each participant $P_i$ selects a $m-length$ random sequence $r_i=(r_{i,1}, r_{i,2}, \ldots, r_{i,m})\in \{0, 1, \ldots , d-1\}^m$, and performs the shift operator $U_{r_{i,j}}(j=1,2, \ldots, m)$ in the form of equation ( \ref{Shift operator} )to the $j-th$ particle of the sequence $S_i$. Then he sends the resulted sequence $\overline{S_i}$ together with $k$ decoy particles (similar to \textbf{Step 1} ) to $TP$.\

\textbf{Step 4 Measurement.}  Having received the sequence from every participant $P_i$, $TP$ will execute eavesdropping checking with every $P_i$ separately similar to \textbf{Step 2.} After successfully passed the eavesdropping checking, $TP$ will extracts $\overline{S_i}$ by deleting the decoy particles. Next, he measures each particle in $\overline{S_i}$ on the Z-basis, and the measurement result is denoted by $|w_i \rangle =|w_{i,1} \rangle |w_{i,2} \rangle \cdots|w_{i,m} \rangle $. \\
\textbf{Step 5 Transmitting privacy.}  Each participant $P_i$ encrypts his privacy $p_i=(p_{i,1}, p_{i,2}, \ldots, p_{i,m})$ into $\overline{p_i}=(\overline{p_{i,1}}, \overline{p_{i,2}}, \ldots, \overline{p_{i,m}})= (p_{i,1}\ominus r_{i,1}, p_{i,2} \ominus r_{i,2}, \ldots, p_{i,m} \ominus r_{i,m})$, and sends it to $TP$ through an authenticated channel.
\textbf{Step 6 Comparison.} Having received $\overline{p_i}$ from every participant $P_i$, $TP$ calculates:
 \begin{equation}\label{Comparison}
 \begin{array}{ll}
 ~~~t_i&=(t_{i,1}, t_{i,2}, \ldots, t_{i,m})\\[2mm]
    &=(\overline{p_{i,1}}\oplus w_{i,1}, \overline{p_{i,2}}\oplus w_{i,2}, \ldots, \overline{p_{i,m}}\oplus w_{i,m})\\[2mm]
 t(i,{i\prime})&=(t_{i,1}\ominus t_{{i\prime},1}, t_{i,2}\ominus t_{{i\prime},2}, \ldots,
                  t_{i,m}\ominus t_{{i\prime},m}))\\[2mm]
 s(i,{i\prime})&=( s(i,{i\prime})_1, s(i,{i\prime})_2, \cdots, s(i,{i\prime})_m )\\[2mm]
       &=(Sign[t_{i,1}\ominus t_{{i\prime},1}], Sign[t_{i,2}\ominus t_{{i\prime},2}], \ldots, Sign[t_{i,m}\ominus t_{{i\prime},m})])\\[2mm]
 \end{array}
 \end{equation}
 where $i,i\prime \in \{0,1,\cdots,k-1\}, i<i\prime$ , $i<j$ and $Sign[\cdot]$ is the signal function which is defined by:
\begin{equation}\label{Signal function}
\begin{array}{l}
Sign[x]=
\left \{
\begin{array}{ll}
  1  & x\in \{1,2,\cdots,l\} \\
  0  & x=0 \\
  -1 & x\in \{l+1,l+2,\cdots,2l\}
\end{array}
\right.
\end{array}
 \end{equation}

 For the $jth(j=1,2,\cdots,m)$ elements of all participants' privacies $p_{0,j}, p_{1,j}, \ldots, p_{k-1,j} $, $TP$ can deduces the size relationship of them from the values of $s(i,{i\prime}) ( i,{i\prime}=0,1,\cdots,k-1 )$. The rules of judgement are as follows:

  \begin{equation}\label{Decision rule}
\begin{array}{l}
 If~~s(i,{i\prime})_j =1, ~~then~~p_{i,j} > p_{{i\prime},j};\\[2mm]
 If~~s(i,{i\prime})_j =0, ~~then~~p_{i,j}=p_{{i\prime},j};\\[2mm]
 If~~s(i,{i\prime})_j =-1, ~~then~~p_{i,j}<p_{{i\prime},j}.
 \end{array}
 \end{equation}

 Next, $TP$ arranges the elements $p_{0,j}, p_{1,j}, \ldots, p_{k-1,j} $ in ascending order together with a relationship symbol $<$ or $=$ between every two elements, and gets a relation expression $p_{i_0^1,j} \lessdot p_{i_1^1,j} \lessdot \ldots \lessdot p_{i_{k-1}^1,j}$, where $i_0^j, i_1^j, $ $\cdots, i_{k-1}^j$ is a permutation of $0, 1, \cdots,$ $ k-1$ , and $\lessdot$ denotes the symbol $<$ or $=$ .

  At last, for each $j\in \{1,2,\cdots,m\}$, $TP$ publishes the information $R_j \triangleq i_0^j \lessdot i_1^j \lessdot \cdots \lessdot i_{k-1}^j $ , which is consisting of subscripts information of the relation expression.  So far, all participants can get the comparison results from $R_j(j\in \{1,2,\cdots,m\})$.

\subsection{Correctness of the protocol}

For the convenience of description,  the phase of eavesdropping checking in step 2 is not considered. Next, we will show that our protocol can work efficiently if all participants and $TP$ execute the protocol honestly. Consider the $jth$ elements $p_{0,j}, p_{1,j}, \ldots, p_{k-1,j} (j=1,2,\cdots,m)$ of all participants.

(a) $TP$ prepares a sequence of  $d-level$ $k-particle$ GHZ states:
  $$|\Phi \rangle _{0,1,\cdots, k-1}=\frac{1}{\sqrt{d}}(|0 \rangle |0 \rangle  \cdots |0 \rangle +|1 \rangle |1 \rangle  \cdots |1 \rangle +\cdots +|d-1 \rangle |d-1 \rangle \cdots|d-1 \rangle )_{0,1,\cdots, k-1}$$
He splits it into $k$ single particle sequence $S_0, S_1, \ldots, S_{k-1}$ and sends the $ith$ sequence $S_i$ to $P_i$.

(b) In the step 3 , each participant $P_i$ selects a $r_i=(r_{i,1}, r_{i,2}, \ldots,$ $ r_{i,m})\in \{0, 1, \ldots , d-1\}^m$ randomly, and performs the shift operator $U_{r_{i,j}}$ in the form of equation ( \ref{Shift operator} )to the $jth$ particle in his own hand. Then he sends the resulted sequence to $TP$.

(c)In the step 4, the final state of the $jth$ GHZ state will be as follows:
\begin{equation}\label{final state}
\begin{array}{rl}
  &|\Phi \rangle _{0,1,\cdots, k-1}\\[2mm]
  =&\frac{1}{\sqrt{d}}(|0\oplus r_{0,j} \rangle |0\oplus r_{1,j} \rangle  \cdots |0\oplus r_{d-1,j} \rangle +|1\oplus r_{0,j} \rangle |1\oplus r_{1,j} \rangle  \cdots |1\oplus r_{d-1,j} \rangle +\\[2mm]
  &\cdots +|(d-1)\oplus r_{0,j} \rangle |(d-1)\oplus r_{1,j} \rangle \cdots |(d-1)\oplus r_{d-1,j} \rangle )_{0,1,\cdots, k-1}\\[2mm]
  =&\frac{1}{\sqrt{d}}(|r_{0,j} \rangle |r_{1,j} \rangle  \cdots |r_{d-1,j} \rangle +|1\oplus r_{0,j} \rangle |1\oplus r_{1,j} \rangle  \cdots |1\oplus r_{d-1,j} \rangle +\cdots +\\[2mm]
  &|(d-1)\oplus r_{0,j} \rangle |(d-1)\oplus r_{1,j} \rangle \cdots |(d-1)\oplus r_{d-1,j} \rangle )_{0,1,\cdots, k-1}
\end{array}
\end{equation}
$TP$ measures it in the Z-basis, and the state will collapse into one of the following states:
\begin{equation}\label{final state 2}
 \begin{array}{l}
 |r_{0,j} \rangle |r_{1,j} \rangle  \cdots |r_{d-1,j} \rangle \\[2mm]
 |1\oplus r_{0,j} \rangle |1\oplus r_{1,j} \rangle  \cdots |1\oplus r_{d-1,j} \rangle \\[2mm] \cdots \\[2mm]
|(d-1)\oplus r_{0,j} \rangle |(d-1)\oplus r_{1,j} \rangle \cdots |(d-1)\oplus r_{d-1,j} \rangle\\[2mm]
 \end{array}
 \end{equation}
Hence, there exists an $c_j \in \{0, 1, \cdots, d-1\}$, the GHZ state in the form of equation (\ref{final state}) will collapse into $|c_j\oplus r_{0,j} \rangle |c_j\oplus r_{1,j} \rangle \cdots |c_j\oplus r_{d-1,j} \rangle $, which implies that $|w_{i,j} \rangle =|c_j\oplus r_{i,j} \rangle  ( i = 0, 1, \cdots, d-1)$.

(d) Each participant $P_i$ encodes his privacy $p_i=(p_{i,1}, p_{i,2}, \ldots, p_{i,m})\in \{0, 1, \ldots , l\}^m$ into $\overline{p_i}=\overline{p_{i,1}}= (p_{i,1}\ominus r_{i,1}, p_{i,2}\ominus r_{i,2}, \ldots, p_{i,m}\ominus r_{i,m})\in \{0, 1, \ldots , l\}^m$, and sends $\overline{p_i}$ to $TP$. Note that the $jth$ elements of all participants' privacies are encoded into $p_{0,j}\ominus r_{0,j}, p_{1,j}\ominus r_{1,j}, \cdots,  p_{m-1,j}\ominus r_{m-1,j} $.

(e) At last, $TP$ calculates the equation(\ref{Comparison}).  Now, we only consider $s(i,{i\prime})_j= Sign[t_{i,j}\ominus t_{{i\prime},j})]$( $ i, i\prime \in \{0, 1, \cdots, d-1\}$)：
\begin{equation}\label{Proof of Comparison}
\begin{array}{rl}
 &t_{i,j}\ominus t_{{i\prime},j}\\[2mm]
=&(\overline{p_{i,j}}\oplus w_{i,j})\ominus (\overline{p_{i\prime,j}}\oplus w_{i\prime,j})\\[2mm]
=&[(p_{i,j}\ominus r_{i,j})\oplus (c_j\oplus r_{i,j})]\ominus [(p_{i\prime,j}\ominus r_{i\prime,j}) \oplus (c_j\oplus r_{i\prime,j})]\\[2mm]
=&p_{i,j}\ominus p_{i\prime,j}
\left \{
\begin{array}{lc}
\in \{1,2,\cdots, l\}     &   p_{i,j}> p_{i\prime,j}\\[2mm]
=0                        &   p_{i,j}= p_{i\prime,j}\\[2mm]
\in \{l+1,l+2,\cdots, l\} &   p_{i,j}< p_{i\prime,j}\\[2mm]
\end{array}
\right.
\end{array}
 \end{equation}
 Then $TP$ will get
\begin{equation}\label{Proof of Sign[]001}
\begin{array}{l}
Sign[t(i,i\prime)]=
\left \{
\begin{array}{lc}
1        &    p_{i,j}> p_{i\prime,j}\\[2mm]
0        &    p_{i,j}= p_{i\prime,j}\\[2mm]
-1       &    p_{i,j}< p_{i\prime,j}\\[2mm]
\end{array}
\right.
\end{array}
 \end{equation}
From $s(i,{i\prime})_j= Sign[t_{i,j}\ominus t_{{i\prime},j})]$($ i, i\prime \in \{0, 1, \cdots, d-1\}$), $TP$ can give the size relationship of $p_{0,j}, p_{1,j}, \ldots, p_{k-1,j} (j=1,2,\cdots,m)$ correctly.

\subsection{A novel example of the protocol}
 Let us give a novel example for illustration without considering the eavesdropping checking. Let $k=3, m=2, l=4$, and $d=2l+1=9$. The privacies of $P_0, P_1$ and $P_2$ are $p_0=(1,4)$, $p_1=(2,2)$, and $p_2=(2,3)$.

 (1) $TP$ prepares  2 identical $9-level$ $3-particle$ GHZ states $|\Phi\rangle^1_{0,1,2}=|\Phi\rangle^2_{0,1,2}=\frac{1}{3}(|0\rangle|0\rangle|0\rangle+|1\rangle|1\rangle|1\rangle+\cdots +|8\rangle|8\rangle|8\rangle)_{0,1,2}$, splits them into 3 particle $S_0, S_1$ and $S_2$ and sends them to $P_0, P_1$ and $P_2$ separately.

 (2) $P_0$ ($P_1$, $P_2$) selects a $2-length$ random sequence $r_0=(4, 6)$ ($r_1=(2, 5)$, $r_2=(6, 1)$), performs the shift operator $U_{r_{0,j}}$ ($U_{r_{1,j}}$, $U_{r_{2,j}}$) to the $j-th$ particle of the sequence $S_0$ ( $S_1$, $S_2$), where $j=1,2$, and sends the resulted particle sequence to $TP$.\\
 (3) At this moment, $TP$ possesses the $3-particle$ GHZ states $|\Phi\rangle^1_{0,1,2}$ and $ |\Phi\rangle^2_{0,1,2}$ which will be\\
\begin{equation}\label{The final state}
\begin{array}{lll}
  |\Phi\rangle^1_{0,1,2} & = &\frac{1}{3}(|0\oplus r_{0,1}\rangle|0\oplus r_{1,1}\rangle|0 \oplus r_{2,1}\rangle+|1\oplus
                        r_{0,1}\rangle|1\oplus r_{1,1}\rangle|1 \oplus r_{2,1}\rangle+\\
                   &   &  \cdots +|8\oplus r_{0,1}\rangle|8\oplus r_{1,1}\rangle|8 \oplus r_{2,1}\rangle)_{0,1,2} \\
                   & = & \frac{1}{3}(|4\rangle|2\rangle|6\rangle+|5\rangle|3\rangle|7\rangle+\cdots +|3\rangle|1\rangle|5\rangle)_{0,1,2} \\
  |\Phi\rangle^2_{0,1,2} & = &\frac{1}{3}(|0\oplus r_{0,2}\rangle|0\oplus r_{1,2}\rangle|0 \oplus r_{2,2}\rangle+|1\oplus
                         r_{0,2}\rangle|1\oplus r_{1,2}\rangle|1 \oplus r_{2,2}\rangle+\\
                   &   &  \cdots +|8\oplus r_{0,2}\rangle|8\oplus r_{1,2}\rangle|8 \oplus r_{2,2}\rangle)_{0,1,2} \\
                   & = & \frac{1}{3}(|6\rangle|5\rangle|1\rangle+|7\rangle|6\rangle|2\rangle+\cdots +|5\rangle|4\rangle|0\rangle)_{0,1,2}
\end{array}
 \end{equation}
 $TP$ measures each particle in $|\Phi\rangle^1_{0,1,2}$ and $|\Phi\rangle^2_{0,1,2}$ on the Z-basis, he will get $|w_0\rangle=|w_{0,1}\rangle|w_{0,2}\rangle=|c_1\oplus r_{0,1}\rangle|c_2\oplus r_{0,2}\rangle$, $|w_1\rangle=|w_{1,1}\rangle|w_{1,2}\rangle=|c_1\oplus r_{1,1}\rangle|c_2\oplus r_{1,2}\rangle$, $|w_2\rangle=|w_{2,1}\rangle|w_{2,2}\rangle=|c_1\oplus r_{2,1}\rangle|c_2\oplus r_{2,2}\rangle$,  where $c_1, c_2 \in \{0, 1, \cdots, 8\}$. For example, if $|w_0\rangle=|4\rangle|7\rangle$, then $c_1=0$, $c_2=1$, $|w_1\rangle=|2\rangle|6\rangle$, $|w_2\rangle=|6\rangle|2\rangle$. \\
 (4) $P_0$ ($P_1$, $P_2$) encodes his privacy into $\overline{p_0}= (p_{0,1}\ominus r_{0,1}, p_{0,2} \ominus r_{0,2})=(6,7)$ (similarly, $\overline{p_1}=(0,6)$, $\overline{p_2}=(5,2)$ ) by $r_0$($r_1$, $r_2$), and sends it to $TP$ through an authenticated channel.\\
 (5) $TP$ calculates:
 $$\begin{array}{l}
 t_0=\overline{p_{0}} \oplus w_{0}=(6,7) \oplus (4,7)=(1,5)\\[2mm]
 t_1=\overline{p_{1}} \oplus w_{1}=(0,6) \oplus (2,6)=(2,3)\\[2mm]
 t_2=\overline{p_{2}} \oplus w_{2}=(5,2) \oplus (6,2)=(2,4)\\[2mm]
 t(0,1)=t_0 \ominus t_1 = (8,2)\\[2mm]
 t(0,2)=t_0 \ominus t_2 = (8,1)\\[2mm]
 t(1,2)=t_1 \ominus t_2 = (0,8)\\[2mm]
 s(0,1)=(Sign[8], Sign[2])=(-1,1)\\[2mm]
 s(0,2)=(Sign[8], Sign[1])=(-1,1)\\[2mm]
 s(1,2)=(Sign[0], Sign[8])=(0,-1)\\[2mm]
\end{array}$$

From the equation(\ref{Decision rule}) and $s(0,1)=(-1,1)$, $TP$ will get $p_{0,1}<p_{1,1}$ and $p_{0,2}>p_{1,2}$. Similarly, $TP$ will get $p_{0,1}<p_{2,1}$ and $p_{0,2}>p_{2,2}$ , $p_{1,1}=p_{2,1}$ and $p_{1,2}<p_{2,2}$. Hence, $TP$ obtains the size relationship of their privacies, i.e., $p_{0,1}<p_{1,1}=p_{2,1}$ and $p_{1,2}<p_{2,2}<p_{0,2}$. At last, he publishes the information $R_1 \triangleq 0 < 1 = 2 $ and $R_2 \triangleq 2 < 3 < 0 $.

 \section{Security analysis and efficiency comparison}
 In this section, we will analyze the security of our protocol from both external and internal attacks. Also, we will analyze the efficiency of our protocol and compare it with other exited protocols.
 \subsection{Security analysis of the protocol}
\textbf{ Case 1 External attack.}
 Suppose that an outsider eavesdropper, Eve, tries to obtain the privacies of participants. From the procession of the protocol, the privacy of each participant $P_i$ is transmitted only once and is encrypted by a random sequence $r_i=(r_{i,0}, r_{i,1}, \cdots, r_{i,m})$. Hence, Eve must find a way to intercept the sequence $r_i=(r_{i,0}, r_{i,1}, \cdots, r_{i,m})$ in Step 3 and the encrypted sequence $\overline{p_i}=(p_{i,1}\ominus r_{i,1}, p_{i,2} \ominus r_{i,2}, \ldots, p_{i,m} \ominus r_{i,m})$ in Step 5. To obtain $r_i$, he must carry out intercept-resend attack, i.e., he intercepts and takes measurements on the particles of $S_i$ and the particles of $\overline{S_i}$, and resents them to receiver.  Let us take the intercept-resend attack on the particles of $S_i$ for example. Due to the existence of the decoy states, Eve need to choose the correct position and measurement-basis of each decoy state in order not to detected by the eavesdropping checking. However, he does not have any information on the position and measurement-basis of each decoy state. If he chooses the right position and right basis, no error will be introduced; or else, the probability of introducing error will be at least $\frac{d-1}{d}$. Hence, his eavesdropping behavior will be detected with $1-(\frac{d-1}{2d})^m$, which will approaches to 1 when $m$ is large enough. It is the same with the case of intercept-resend attack on the the particles of $\overline{S_i}$. Therefore, Eve can not obtain the random sequence $r_i=(r_{i,0}, r_{i,1}, \cdots, r_{i,m})$. Also, he can not obtain the sequence $\overline{p_i}=(p_{i,1}\ominus r_{i,1}, p_{i,2} \ominus r_{i,2}, \ldots, p_{i,m} \ominus r_{i,m})$ in Step 5 because the channels between the  $TP$ and participants are authenticated. From the analysis above, the protocol is immune to external attack.\\
\textbf{ Case 2 Internal attack from participants.}
Suppose that a participant, $P_0$, is a dishonest participant who tries to obtain the privacies of other participants, and $TP$ is the semi-honest party who will not collude with anyone. If $P_0$ wants to steal the privacy of a certain participant $P_i(i\in \{1,2,\cdots,d-1\})$, he could firstly measures the particles in the sequence of $S_0$ on the Z-basis before performing the random shift operators on them, and the measurement results are identical to the particles in $S_i$. Next, to obtain the random sequence $r_i$, $P_0$ needs to measure the particles in the sequence $\overline{S_i}$ by using the intercept-resend attack. In this environment, $P_0$ can be considered as an outside attacker, and his interception behavior will be caught by $P_i$ and $TP$ similar to the case of external attack. Also, $P_0$ can not obtain the sequence $\overline{p_i}$ in Step 5 because the channel between  $TP$ and $P_i$ is authenticated. The collusion attack from multiple participant is the same.\\
\textbf{ Case 3 Internal attack from the semi-honest third party $TP$.}  Obviously, the dishonest third party $TP$ is the one who can get the most information during the execution of the protocol. However, due to his semi-honesty, he will prepare the $k-particle$ $d-level$ GHZ states rather than other types of particles such as single particles(even if he prepared other quantum states, his dishonest behavior would be discovered by participants in the following way. Before step 3, all participants consult to select some positions of particles randomly, and  measure each particle of these positions using either X-basis or Z-basis. They can verify whether these quantum states are GHZ states or not by publishing the measurement results). Next, $TP$ will execute the protocol honestly. The only way to derive the privacy of $P_i$ relies on the analysis of information received from $P_i$. Firstly, he can obtain $\overline{p_i}= (p_{i,1}\ominus r_{i,1}, p_{i,2} \ominus r_{i,2}, \ldots, p_{i,m} \ominus r_{i,m})$ legally in Step 5. So he needs to get the random sequence $r_i=(r_{i,1}, r_{i,2}, \ldots, r_{i,m})\in \{0, 1, \ldots , d-1\}^m$ and nextly extracts the privacy of $P_i$. Apparently, the random sequence $r_i=(r_{i,1}, r_{i,2}, \ldots, r_{i,m})\in \{0, 1, \ldots , d-1\}^m$ is encoded into the sequence $\overline{S_i}$ which is entangle with $\overline{S_j}$s. When it comes to measure the particles in the sequence $\overline{S_i}$, $TP$ will randomly get one of the following states: $|r_{i,1}\rangle|r_{i,2}\rangle\cdots |r_{i,m}\rangle$, $|r_{i,1}\oplus 1\rangle|r_{i,2}\oplus 1\rangle\cdots |r_{i,m}\oplus 1\rangle, \cdots$, $|r_{i,1}\oplus (d-2)\rangle|r_{i,2}\oplus (d-2)\rangle\cdots |r_{i,m}\oplus (d-2)\rangle$ and $|r_{i,1}\oplus (d-1)\rangle|r_{i,2}\oplus (d-1)\rangle\cdots |r_{i,m}\oplus (d-1)\rangle$. Hence, $TP$ can not obtain $r_i=(r_{i,1}, r_{i,2}, \ldots, r_{i,m})\in \{0, 1, \ldots , d-1\}^m$ accurately, and can not derive the privacy of $P_i$.\\
\subsection{Efficiency comparison with existed protocols}
Here, we will compare the protocol with four existed MQPC protocols in the following five aspects: quantum resources used, the  category of MQPC (size or equality comparison), the qubit or qudit efficiency which is defined as $\eta = \frac{c}{q+b}$ (here $c$ is the length of privacies of participants, $q$  and $b$ are the numbers of qudits and classical bits used in transmission and eavesdropping checking, whether participants need to share privacy common key beforehand, and security . For the sake of discussion, it is assumed that the length of the privacies is $m$, and the number of decoy particles is equal to the number of quantum particles transmitted in each MQPC protocol. The four existing MQPC protocols are CTH2013 protocol \cite{CTH2013}, HHH2017 protocol \cite{HHH2017}, LYS2014 protocol \cite{LYS2014}, and HHG2015 protocol \cite{HHG2015}. Now, we will show the comparison result as follows ( see table  \ref{Efficiency comparison table} ).\\
\begin{table}
\renewcommand{\arraystretch}{1.9}
\setlength{\abovecaptionskip}{2pt}
\setlength{\belowcaptionskip}{6 pt}
 \centering
 \caption{Comparison between the existed four QPC protocols with ours }\label{Efficiency comparison table}
\begin{tabular}{cccccc}
\hline
QPC Protocol &\makecell[c]{quantum\\resources}  & \makecell[c]{Category\\ of QPC}  & Efficiency $\eta$ & \makecell[c]{Need to share \\privacy key}  & Security \\
 \hline
  CTH2013\cite{CTH2013}&\makecell[c]{$2-level$ GHZ\\ class states}&Equality& $\frac{1}{3k}$       &  No &Secure\\
  HHH2017\cite{HHH2017}&\makecell[c]{$2-level$ Bell\\ states}     &Equality& $\frac{1}{8k}$       &  No &Secure\\
  LYS2014 protocol \cite{LYS2014}&\makecell[c]{$d-level$ entangled\\states}      &Size    & $\frac{1}{3k}$ & Yes  &secure\\
  HHG2015 protocol \cite{HHG2015}&\makecell[c]{$d-level$ GHZ and\\ entangled particles} &Size    & $\frac{1}{6k}$ & No  &Insecure\\
  Ours   &\makecell[c]{$d-level$ GHZ\\states}      &Size    &  $\frac{1}{3k}$      &  No  &Secure\\
\hline
\end{tabular}
\end{table}
(1) CTH2013 protocol. The authors proposed a 4-party QPC protocol, and a multiparty (say $k-party$ hereafter) QPC protocol which are used to compare the equality of the privacies. We only consider the case of $k-party$. The quantum resources used in this protocol are $2-level$ $k-particle$ GHZ-class states. The transmission of information includes two stages. First, $TP$ prepares $m$ $k-particle$ GHZ-class states. Then, he splits them into $k$ particle-sequence and sends every sequence to the corresponding participant with $m$ decoy particles. Second, each participant sends his encoded privacy which is $m$ bits to $TP$. Hence, the efficiency $\eta = \frac{m}{mk+mk+mk}=\frac{1}{3k}$. Besides, the participants need not to share privacy common key beforehand, and the protocol is secure at present because there is no efficient attack for it.\\
(2) HHH2017 protocol. The authors proposed a $k-party$ QPC protocol of comparing the equality in which two $TP$s are introduced to deal with the comparison in a strange environment. The quantum resources used in this protocol are $2-level$ $k-particle$ GHZ-class states. The transmission of information includes three stages. First, $TP_1$  prepares $2m$ $k-particle$ GHZ-class states. Then, he splits them into $k$ particle-sequence and sends every sequence to the corresponding participant with $2m$ decoy particles. Second, $TP_1$ sends the information of the GHZ states to $TP_2$ using quantum secure direct communication and the quantum resource used here is at least $2mk$ qubits. Third, each participant sends his encoded privacy which is $m$ bits to $TP_1$ and $TP_2$. Hence, the efficiency $\eta = \frac{m}{2mk+2mk+2mk+mk+mk}=\frac{1}{8k}$. Besides, the participants also need not to share privacy common key beforehand, and the protocol is secure at present.\\
(3) LYS2014 protocol. The authors proposed a $k-party$ QPC protocol of comparing the sizes of privacies. The quantum resources used in this protocol are $d-level$ entangled states, and the participants need to share a privacy common key $K$ beforehand through a secure QKA protocol. The transmission of information contains three step. First, $TP$ prepares $m$ $k-particle$ $d-level$ entangled states. Then, he splits them into $k$ particle-sequence and sends each sequence to the corresponding participant with $m$ decoy particles. Second, each participant measures the received particle-sequence which will be transformed into a classical $m-bit$ sequence, and he encrypts his privacy by the classical bit-sequence and the privacy common key $K$ using one-time pad. At last, each participant sends his encrypted privacy information($m-bit$ sequence) to $TP$ through an authenticated channel. Hence, the efficiency $\eta = \frac{m}{mk+mk+mk}=\frac{1}{3k}$. However, the actual efficiency is lower than $\frac{1}{3k}$ because the participants need to share a privacy common key $K$ beforehand through a QKA protocol which will waste a lot of quantum resource. This protocol is secure at present because there is no efficient attack for it.\\
(4)HHG2015 protocol. The authors proposed a $k-party$ QPC protocol of comparing the sizes of privacies. The quantum resources used in this protocol are $d-level$ GHZ states and $d-level$ entangled states. The transmission of information includes two stages. First, $TP$ prepares $m$ $k-particle$ $d-level$ GHZ states and $m$ $k-particle$ $d-level$ entangled states. Then, $TP$ splits the $m$ $k-particle$ $d-level$ GHZ states into $k$ particle-sequences and sends them to the corresponding participant with $m$ decoy particles. Also, he splits the $m$ $k-particle$ $d-level$ entangled states into $k$ particle-sequences and sends them to the corresponding participant with $m$ decoy particles. Second, each participant measures the first $k-particle$ sequence. Then he performs the unitary operations, which are decided by the measurement results and his privacy, on the second $k-particle$ sequence and sends the resulted $k-particle$ sequence with $m$ decoy particles to $TP$. The efficiency $\eta = \frac{m}{2mk+2mk+mk+mk}=\frac{1}{6k}$. Besides, the participants need not to share a privacy common key beforehand. Hence, the HHG2015 protocol is much more efficient than LYS2014 protocol. \\
However, there is a serious bug in the HHG2015 protocol. From step 4 and step 6, We can easily get that $p_i=p_j$ and $q_i=q_j$  for each $i$ and $j$. If a dishonest participant(say $P_1$), wants to steal the privacy of another one(say $P_2$), he will firstly intercept the particles sent from $P_2$ and resents forged particles to $TP$ in step 4. Secondly, he deletes the decoy particles and measures the remaining particles after $P_2$ published the positions of decoy states. Therefore, $P_1$ will get the value of $MR_2 = (s_2+p_2+q_2)$mod$d$ and $s_2 = (MR_2-p_2-q_2)$mod$d$ which is the privacy of $P_2$. Although this attack will be discovered by $TP$ and $P_2$, but they did not know the identity of the attacker. Hence, $P_1$  succeeded in obtaining the privacy of $P_2$. Similarly, he can get the privacy of any other participant without being found. So, this protocol is insecure. \\
(5) Our protocol. We proposed a $k-party$ QPC protocol of comparing the sizes of privacies by using  $k-particle$ $d-level$ GHZ states, and the participants need not to share privacy common key beforehand. The transmission of information includes three stages. First, $TP$ prepares $m$ $k-particle$ GHZ-class states. Then, he splits them into $k$ particle-sequence and sends each sequence to the corresponding participant with $m$ decoy particles. Second, After encoding the received sequence by a series of random unitary operations, each participant inserts $m$ decoy particles into it and sends it back to $TP$. Third, every participant transmits $m$ classical bits to $TP$ separately. Hence, the efficiency $\eta = \frac{m}{mk+mk+mk}=\frac{1}{3k}$ which is as good as that of the LYS2014 protocol and CTH2013 protocol, and our protocol is secure  against external and internal attacks. However, owing to the waste of quantum resource in the sharing the common key beforehand through a QKA protocol in the LYS2014 protocol, our protocol is more efficient than it because  the participants need not to share private common key beforehand in our protocol. Besides, the CTH2013 protocol only solves the problem of equality comparison. Therefore, our protocol is better than the LYS2014 protocol and CTH2013 protocol.\\
\section{Conclusion}
We presented a MQPC protocol with $k-particle$ $d-level$ GHZ states. In the protocol, all participants can compare the size of their privacy with the help of a semi-honest party $TP$. Besides, we gave a novel example of the proposed protocol. Security analysis shows that it is immune to both external attack and internal attack in theory, and efficiency comparison shows that it is prior to all existing protocols of the same type. However, our protocol is only suitable for scenarios in an ideal environment. How to improve the agreement to adapt to a more complicated environment is our main work in the future. \\
\section*{Acknowledgments}
This work was supported in part by the National Key R\&D Program of China under Grant 2017YFB0802400, the National Science Foundation of China under Grant 61373171, 61702007 and 11801564, the 111 Project under Grant B08038, and the Key Project of Science Research of Anhui Province under Grant KJ2017A519, the Natural Science Foundation of Shaanxi Province under Grant No.2017JQ1032, the Basic Research Project of Natural Science of Shaanxi Province under Grant 2017JM6037.

\end{spacing}

\end{document}